 \documentclass[final,5p,times,twocolumn,sort&compress]{elsarticle}

\usepackage [latin1]{inputenc}
\usepackage{amssymb,bm}
\usepackage{lipsum}
\usepackage{amsmath}
\usepackage{hyperref}
\usepackage{microtype}

\newcommand{\I}{\mathbb{I}}

\journal{Physics Letters B}

\begin{document}

\begin{frontmatter}



\title{On-shell reduction of spin-zero Duffin--Kemmer--Petiau covariants and implications for the relativistic impulse approximation}


\author[first,second]{Luis B. Castro}\ead{lrb.castro@ufma.br}
\affiliation[first]{organization={Programa de P\'{o}s-gradua\c{c}\~{a}o em F\'{i}sica, Universidade Federal do Maranh\~{a}o},
            addressline={Campus Universit\'{a}rio do Bacanga}, 
            city={S\~{a}o Lu\'{i}s},
            postcode={65080-805}, 
            state={MA},
            country={Brazil}}
\affiliation[second]{organization={Coordena\c{c}\~{a}o do Curso de F\'{i}sica - Bacharelado, Universidade Federal do Maranh\~{a}o},
            addressline={Campus Universit\'{a}rio do Bacanga}, 
            city={S\~{a}o Lu\'{i}s},
            postcode={65080-805}, 
            state={MA},
            country={Brazil}}

\begin{abstract}

We analyze the on-shell content of the local Lorentz covariants of the five-dimensional spin-zero Duffin--Kemmer--Petiau algebra in elastic spin-zero--spin-$1/2$ scattering. The scalar, vector, and tensor structures are reduced explicitly to the two independent amplitudes allowed by the physical parity-conserving process. The two scalar structures are proportional on shell, the genuine nonminimal vector $i[P,\beta^\mu]$ is longitudinal and vanishes after contraction with the conserved fermion vector current, and the projected vectors $P\beta^\mu$ and $\beta^\mu P$ each reduce to one half of the minimal-vector contribution. The tensor covariants introduce no additional on-shell directions. We discuss the consequences for the four invariant choices used in meson--nucleus optical-potential calculations and show that different nuclear equations can arise only after on-shell-equivalent representatives are promoted to optical operators and iterated beyond the order at which the elementary impulse approximation is fixed. These results provide a systematic basis for distinguishing genuinely independent elementary amplitudes from representation-dependent effects generated by the optical embedding.
\end{abstract}




\end{frontmatter}




\section{Introduction}
\label{sec:introduction}

The Duffin--Kemmer--Petiau (DKP) equation provides a first-order relativistic description of spin-zero and spin-one bosons \cite{Petiau1936,Kemmer1938,PR54:1114:1938,Kemmer1939}. In the irreducible spin-zero representation, the bosonic field is described by a five-component spinor acted on by $5\times5$ DKP matrices~\cite{Umezawa1956,PLA268:165:2000}. The formalism provides an alternative to the second-order Klein--Gordon description and has found an important application in relativistic descriptions of meson--nucleus scattering~\cite{PRL55:592:1985,PLB427:231:1998,PTP103:321:2000}.

A historically important application of the five-dimensional spin-zero DKP formalism is the relativistic impulse approximation (RIA) for meson--nucleus scattering introduced by Clark \emph{et al.} \cite{PRL55:592:1985}. In that construction, empirical spin-zero--nucleon amplitudes are converted into scalar and vector invariants and folded with relativistic scalar and vector nuclear densities. Several choices of scalar and vector DKP operators were considered, giving rise to four cases that were subsequently used in pion- and kaon--nucleus calculations \cite{PRL55:592:1985,PLB427:231:1998,PTP103:321:2000}.

The multiplicity of these choices reflects the enlarged operator content of the spin-zero DKP algebra. The $25$ linearly independent matrices of the five-dimensional representation can be organized into two scalar, two vector, and two second-rank tensor Lorentz structures \cite{PRD15:1518:1977,AP104:427:1977}. The projected vectors $P\beta^\mu$ and $\beta^\mu P$ used in the historical KDP-RIA are therefore only particular DKP representatives. In contrast, the genuine nonminimal vector coupling is associated with $i[P,\beta^\mu]$ \cite{JPA43:055306:2010,EPJC77:310:2017}.

The elementary elastic process relevant to the first-order RIA is, however, much more constrained than this operator basis might suggest. For parity-conserving scattering of a spin-zero boson from a spin-$1/2$ fermion, the on-shell amplitude contains only two independent invariant functions \cite{NPA668:97:2000}. The central question is therefore not how many matrices occur in the DKP algebra, but how the complete set of local DKP Lorentz covariants projects onto this two-dimensional physical amplitude space.

The purpose of this Letter is not to reestablish the two-function structure of the physical amplitude, but to determine explicitly how the scalar, vector, and tensor DKP covariants represent those two on-shell directions. We first reduce the complete local spin-zero DKP covariant basis between free on-shell boson states and then contract the resulting structures with the corresponding on-shell fermion bilinears. We subsequently apply the resulting identities to the four invariant choices used in meson--nucleus optical-potential calculations~\cite{PRL55:592:1985,PLB427:231:1998,PTP103:321:2000} and to their optical embedding. The central result is that the enlarged DKP covariant basis does not generate additional elementary on-shell amplitudes; differences between covariant choices can instead emerge after on-shell-equivalent representatives are promoted to optical operators.

This Letter is organized as follows. Section~\ref{sec:scattering-covariants} establishes the physical two-amplitude structure and the complete local spin-zero DKP covariant basis. Section~\ref{sec:onshell-reduction} performs the on-shell reduction, Sec.~\ref{sec:ria-implications} applies the resulting identities to the historical KDP-RIA and its optical embedding, and Sec.~\ref{sec:discussion} summarizes the physical implications and conclusions.

\section{On-shell spin-zero--spin-$1/2$ scattering and DKP covariants}
\label{sec:scattering-covariants}

The elementary meson--nucleon amplitude entering the relativistic impulse approximation (RIA) must connect the physical spin-zero--spin-$1/2$ scattering amplitude with the DKP operator structures used to represent the boson \cite{PRL55:592:1985}.  We therefore first establish the on-shell kinematics and the two-function structure of this amplitude, and then specify the local Lorentz structures available in the five-dimensional spin-zero DKP algebra.  This separation keeps the physical constraints of the scattering process distinct from the particular DKP covariants used to represent it.

We consider elastic scattering of a spin-zero boson $\phi$ of mass $m$ from a spin-$1/2$ fermion $N$ of mass $M$.  The notation $\phi(k)$ and $\phi(k')$ labels the incoming and outgoing boson states with four-momenta $k^\mu$ and $k'^\mu$, while $N(p)$ and $N(p')$ label the corresponding fermion states with four-momenta $p^\mu$ and $p'^\mu$.  The elementary process is
\begin{equation}
 \phi(k)+N(p)\longrightarrow \phi(k')+N(p').
 \label{eq:process}
\end{equation}
In the meson--nucleon applications motivating the RIA, $\phi$ denotes a pion or kaon and $N$ a nucleon \cite{PRL55:592:1985}.  Isospin labels are suppressed because they do not affect the Lorentz structure considered here.  We use natural units, $\hbar=c=1$, and the metric $g^{\mu\nu}=\mathrm{diag}(1,-1,-1,-1)$.

Because the states in Eq.~\eqref{eq:process} are asymptotic and free, their four-momenta satisfy the on-shell conditions
\begin{equation}
 k^2=k'^2=m^2,
 \qquad
 p^2=p'^2=M^2.
 \label{eq:onshell}
\end{equation}
In addition to the on-shell constraints, elastic scattering conserves the total four-momentum,
\begin{equation}
 k+p=k'+p'.
 \label{eq:fourmomentumconservation}
\end{equation}
These relations provide the kinematic setting for the amplitude below.  We introduce the momentum transfer $q^\mu$, the average boson momentum $\bar k^\mu$, and the average fermion momentum $\bar p^\mu$ through
\begin{equation}
 q^\mu=k'^\mu-k^\mu=p^\mu-p'^\mu,
 \quad
 \bar k^\mu=\frac{k'^\mu+k^\mu}{2},
 \quad
 \bar p^\mu=\frac{p'^\mu+p^\mu}{2}.
 \label{eq:kinematics}
\end{equation}
Together with the on-shell conditions, these definitions imply that both average momenta are orthogonal to the momentum transfer,
\begin{equation}
 \bar k\cdot q=0,
 \qquad
 \bar p\cdot q=0.
 \label{eq:orthogonality}
\end{equation}

In the center-of-mass frame, rotational invariance and parity conservation restrict the elastic spin-zero--spin-$1/2$ amplitude to two independent functions.  Following the convention used in the KDP-RIA literature \cite{PRL55:592:1985}, we write
\begin{equation}
 F(\theta)=f(\theta)+\bm\sigma\cdot\hat{\bm n}\,g(\theta).
 \label{eq:fg}
\end{equation}
Here $f(\theta)$ and $g(\theta)$ are the spin-nonflip and spin-flip amplitudes, respectively, $\bm\sigma$ is the vector of Pauli matrices acting on the fermion spin space, and $\hat{\bm n}=(\bm k\times\bm k')/|\bm k\times\bm k'|$ is the unit vector normal to the scattering plane.

This amplitude can equivalently be represented in a manifestly Lorentz-covariant form.  Denoting the invariant matrix element of the process by $\mathcal M$, one may write \cite{NPA668:97:2000}
\begin{equation}
 \mathcal M=
 \bar u(p')\left[A(s,t)+\!\not\!\bar k\,B(s,t)\right]u(p).
 \label{eq:AB}
\end{equation}
Here $u(p)$ is the free Dirac spinor of the fermion, $\bar u=u^\dagger\gamma^0$, and $\!\not\!\bar k\equiv\gamma_\mu\bar k^\mu$.  The scalar functions $A(s,t)$ and $B(s,t)$ are invariant amplitudes depending on the Mandelstam variables $s=(p+k)^2$ and $t=q^2$.  Equations~\eqref{eq:fg} and \eqref{eq:AB} are therefore two representations of the same physical two-function amplitude.

To represent the bosonic sector of this amplitude in the DKP formalism, we first introduce the free spin-zero state and the corresponding operator algebra.  In the spin-zero DKP representation, the boson state is described by a five-component spinor $\Phi(k)$ satisfying the free DKP equation \cite{Umezawa1956,PLA268:165:2000}
\begin{equation}
 (\beta^\mu k_\mu-m)\Phi(k)=0.
 \label{eq:freeDKP}
\end{equation}
The matrices $\beta^\mu$ entering Eq.~\eqref{eq:freeDKP} are defined by the DKP algebra \cite{Umezawa1956,PLA268:165:2000}
\begin{equation}
 \beta^\mu\beta^\nu\beta^\lambda
 +\beta^\lambda\beta^\nu\beta^\mu
 =g^{\mu\nu}\beta^\lambda+g^{\lambda\nu}\beta^\mu.
 \label{eq:dkpalgebra}
\end{equation}
For the irreducible spin-zero representation, the $\beta^\mu$ are $5\times5$ matrices, so their matrix algebra contains $25$ linearly independent elements.  Under Lorentz transformations these elements can be organized into six types: two scalars, two four-vectors, one antisymmetric second-rank tensor, and one symmetric second-rank tensor \cite{PRD15:1518:1977,AP104:427:1977}.

An explicit realization of these classes can be constructed using the Umezawa projectors.  In the spin-zero representation we define \cite{Umezawa1956,JPA43:055306:2010}
\begin{equation}
 P=\frac{1}{3}\left(\beta^\mu\beta_\mu-\I\right),
 \qquad
 P^\mu=P\beta^\mu.
 \label{eq:Pdef}
\end{equation}
Here $P$ is a scalar projector and $P^\mu$ transforms as a four-vector.  As a consequence of the DKP algebra, these operators satisfy \cite{Umezawa1956,PLA268:165:2000}
\begin{equation}
 P^2=P,
 \qquad
 P^\mu\beta^\nu=Pg^{\mu\nu},
 \qquad
 \{P,\beta^\mu\}=\beta^\mu.
 \label{eq:projectoridentities}
\end{equation}
The first expression in Eq.~\eqref{eq:Pdef} is equivalent, within the irreducible spin-zero representation, to the usual Umezawa product form $-(\beta^0)^2(\beta^1)^2(\beta^2)^2(\beta^3)^2$.

With these definitions, a convenient choice of representatives for the six Lorentz classes is
\begin{align}
 &S_1=\I,\quad S_2=P,\quad V_1^\mu=\beta^\mu,\quad V_2^\mu=i[P,\beta^\mu],\nonumber\\
 &T_A^{\mu\nu}=i[\beta^\mu,\beta^\nu],\quad T_S^{\mu\nu}=\{\beta^\mu,\beta^\nu\}
 -\frac{1}{2}g^{\mu\nu}\beta^\alpha\beta_\alpha.
 \label{eq:completebasis}
\end{align}
The operators $S_1$ and $S_2$ are scalar representatives, $V_1^\mu$ and $V_2^\mu$ are four-vector representatives, $T_A^{\mu\nu}$ is antisymmetric, and $T_S^{\mu\nu}$ is symmetric and traceless.  Accordingly, the two tensors contain six and nine independent components, respectively, and the count $1+1+4+4+6+9=25$ exhausts the five-dimensional matrix algebra.

Equation~\eqref{eq:completebasis} classifies the bosonic DKP sector only.  Lorentz-scalar meson--fermion invariants are obtained by contracting the open indices with fermion bilinears and, when required, external four-momenta.

The projected vector structures used in the historical KDP-RIA can now be expressed in terms of this basis.  Clark and collaborators employed structures containing $P\beta^\mu$ or $\beta^\mu P$ \cite{PRL55:592:1985,PTP103:321:2000}.  The last identity in Eq.~\eqref{eq:projectoridentities} gives
\begin{equation}
 P\beta^\mu=\frac{1}{2}\left(\beta^\mu+[P,\beta^\mu]\right),
 \qquad
 \beta^\mu P=\frac{1}{2}\left(\beta^\mu-[P,\beta^\mu]\right).
 \label{eq:projecteddecomp_setup}
\end{equation}
Thus the historical projected vectors do not define additional Lorentz classes; they are linear combinations of the two vector structures already contained in Eq.~\eqref{eq:completebasis}.

\section{On-shell reduction of the DKP structures}
\label{sec:onshell-reduction}

The six Lorentz classes in Eq.~\eqref{eq:completebasis} span the local spin-zero DKP matrix algebra, but the elementary on-shell amplitude in Eq.~\eqref{eq:AB} contains only two physical invariant functions.  We first reduce all six DKP covariants between free external meson states.  Once this bosonic projection is complete, we contract the resulting structures with on-shell fermion bilinears and determine whether the full meson--fermion amplitude contains any direction beyond the two displayed in Eq.~\eqref{eq:AB}.

Matrix elements between DKP states require the adjoint $\bar\Phi=\Phi^\dagger\eta^0$, with $\eta^0=2(\beta^0)^2-\I$ \cite{Umezawa1956,PLA268:165:2000}.  The free equation in Eq.~\eqref{eq:freeDKP} then has the adjoint form $\bar\Phi(k')(\beta\cdot k'-m)=0$.  Because the projector identities in Eq.~\eqref{eq:projectoridentities} close on $P$, the bilinear $\bar\Phi(k')P\Phi(k)$ appears naturally as a common scalar factor in the on-shell reduction.  We therefore define
\begin{equation}
 \mathcal S\equiv\bar\Phi(k')P\Phi(k).
 \label{eq:Sdef}
\end{equation}
Acting with $P\beta^\mu$ on the free DKP equation and using $P\beta^\mu\beta^\nu=Pg^{\mu\nu}$ gives the first relation below; the corresponding adjoint relation for the outgoing state gives the second:
\begin{equation}
 P\beta^\mu\Phi(k)=\frac{k^\mu}{m}P\Phi(k),
 \qquad
 \bar\Phi(k')\beta^\mu P=\frac{k'^\mu}{m}\bar\Phi(k')P.
 \label{eq:projectedstates}
\end{equation}
Writing $\bar\Phi'\equiv\bar\Phi(k')$ and $\Phi\equiv\Phi(k)$ for brevity, we next use $\beta^\mu=P\beta^\mu+\beta^\mu P$ inside the bilinear $\bar\Phi'\beta^\mu\beta^\nu\Phi$.  The first term is fixed directly by $P\beta^\mu\beta^\nu=Pg^{\mu\nu}$, while the second is reduced with the two relations in Eq.~\eqref{eq:projectedstates}.  This gives
\begin{equation}
 \bar\Phi'\beta^\mu\beta^\nu\Phi
 =\left(g^{\mu\nu}+\frac{k'^\mu k^\nu}{m^2}\right)\mathcal S.
 \label{eq:masterbeta2}
\end{equation}
Equation~\eqref{eq:masterbeta2} is the common starting point for the scalar and tensor reductions, whereas Eq.~\eqref{eq:projectedstates} already determines the vector sector.

We first consider the two scalar representatives.  Applying the free equations once more at both external legs and then using Eq.~\eqref{eq:masterbeta2} gives
\begin{equation}
\bar\Phi'\Phi=\frac{k'_\mu k_\nu}{m^2}
   \bar\Phi'\beta^\mu\beta^\nu\Phi=\left(1+\frac{k'\cdot k}{m^2}\right)\mathcal S
 =2\left(1-\frac{t}{4m^2}\right)\mathcal S.
 \label{eq:scalarreduction}
\end{equation}
Thus $\I$ and $P$ project onto the same scalar direction on shell, with a proportionality factor that depends on the momentum transfer.

The vector classes follow directly from Eq.~\eqref{eq:projectedstates}.  Adding and subtracting the two projected matrix elements according to $\{P,\beta^\mu\}=\beta^\mu$ gives
\begin{equation}
 \bar\Phi'\beta^\mu\Phi
 =\frac{2\bar k^\mu}{m}\mathcal S,
 \qquad
 \bar\Phi' i[P,\beta^\mu]\Phi
 =-\frac{i q^\mu}{m}\mathcal S.
 \label{eq:vectorreduction}
\end{equation}
The minimal vector therefore follows the average meson momentum, whereas the genuine nonminimal vector is purely longitudinal in the transferred momentum.

It remains on the meson side to reduce the two tensor classes.  The antisymmetric representative follows directly from the antisymmetric part of Eq.~\eqref{eq:masterbeta2}.  For the symmetric traceless representative, we take the symmetric part and subtract the trace term required by its definition in Eq.~\eqref{eq:completebasis}.  The two complete tensor matrix elements are therefore
\begin{align}
 \bar\Phi'T_S^{\mu\nu}\Phi
 &=\frac{2\mathcal S}{m^2}\left[
 \bar k^\mu\bar k^\nu-\frac{1}{4}q^\mu q^\nu
 -\frac{1}{4}g^{\mu\nu}\left(m^2-\frac{t}{2}\right)
 \right],
 \label{eq:symtensortraceless}\\
 \bar\Phi'T_A^{\mu\nu}\Phi
 &=\frac{i\mathcal S}{m^2}
 \left(\bar k^\nu q^\mu-\bar k^\mu q^\nu\right).
 \label{eq:antitensor}
\end{align}
Equations~\eqref{eq:scalarreduction}, \eqref{eq:vectorreduction}, \eqref{eq:symtensortraceless}, and \eqref{eq:antitensor} complete the meson-side projection of all six DKP classes.  The resulting tensors contain only the already established vectors $\bar k^\mu$ and $q^\mu$, together with the metric.  This bosonic reduction alone, however, does not yet show that the complete meson--fermion amplitude is restricted to two physical directions.

The remaining step is to contract these projected DKP structures with on-shell fermion bilinears.  Equation~\eqref{eq:AB} shows that the two independent Dirac matrix elements may be represented by
\begin{equation}
 \mathcal U\equiv\bar u(p')u(p),
 \qquad
 \mathcal K\equiv\bar u(p')\!\not\!\bar k\,u(p).
 \label{eq:UKdef}
\end{equation}
We now use $\mathcal U$ and $\mathcal K$ as the physical basis against which the scalar, vector, and tensor sectors are reduced.

For the scalar channel there are no open Lorentz indices to contract.  Multiplying the two DKP scalar matrix elements by the fermion scalar bilinear $\mathcal U$ gives
\begin{equation}
 (\bar\Phi'P\Phi)\,\mathcal U=\mathcal S\mathcal U,
 \qquad
 (\bar\Phi'\Phi)\,\mathcal U
 =2\left(1-\frac{t}{4m^2}\right)\mathcal S\mathcal U.
 \label{eq:scalarfermionclosure}
\end{equation}
Thus both DKP scalar representatives feed the same physical scalar direction $\mathcal U$; their difference is entirely contained in the kinematic coefficient.

For the vector channel, the fermion bilinear carrying the open Lorentz index is the on-shell vector current
\begin{equation}
 J_N^\mu\equiv\bar u(p')\gamma^\mu u(p).
 \label{eq:JNdef}
\end{equation}
The free Dirac equations and $q=p-p'$ imply
\begin{equation}
 q_\mu J_N^\mu
 =\bar u(p')(\!\not\!p-\!\not\!p')u(p)=0.
 \label{eq:currentcons}
\end{equation}
Using $\bar k_\mu J_N^\mu=\mathcal K$, which follows directly from Eqs.~\eqref{eq:UKdef} and \eqref{eq:JNdef}, together with Eq.~\eqref{eq:currentcons}, contraction of the DKP vector matrix elements in Eq.~\eqref{eq:vectorreduction} with $J_N^\mu$ gives
\begin{equation}
 J_{N\mu}\bar\Phi'\beta^\mu\Phi
 =\frac{2\mathcal S}{m}\mathcal K,
 \qquad
 J_{N\mu}\bar\Phi'i[P,\beta^\mu]\Phi=0.
 \label{eq:nmzero}
\end{equation}
The minimal vector supplies the second physical direction $\mathcal K$, whereas the genuine nonminimal vector is invisible in this conserved free on-shell vector-current channel.  The historical projected vectors in Eq.~\eqref{eq:projecteddecomp_setup} then satisfy
\begin{equation}
 J_{N\mu}\bar\Phi'P\beta^\mu\Phi
 =J_{N\mu}\bar\Phi'\beta^\mu P\Phi
 =\frac{1}{2}J_{N\mu}\bar\Phi'\beta^\mu\Phi.
 \label{eq:projectedequiv}
\end{equation}
Thus $P\beta^\mu$ and $\beta^\mu P$ are distinct DKP operators, but they carry the same elementary on-shell vector information after contraction with the conserved fermion current.

The tensor channel is the remaining possible source of additional spin dependence because its two open Lorentz indices allow contractions with both vector and Dirac-tensor bilinears.  The first fermion-side relation needed below follows directly from the free Dirac equations and the definition of $\bar p^\mu$:
\begin{equation}
\bar p_\mu J_N^\mu=\frac{1}{2}\bar u(p')\left(\!\not\!p+\!\not\!p'\right)u(p)=M\bar u(p')u(p)=M\mathcal U.
\label{eq:pJrelation}
\end{equation}
The antisymmetric DKP tensor in Eq.~\eqref{eq:antitensor} is proportional to $\bar k^\nu q^\mu-\bar k^\mu q^\nu$.  Its contraction with a Dirac tensor therefore necessarily produces the combination $\bar k_\mu\bar u(p')i\sigma^{\mu\nu}q_\nu u(p)$.  With $\sigma^{\mu\nu}=\tfrac{i}{2}[\gamma^\mu,\gamma^\nu]$, this is precisely the combination reduced by the Gordon identity, which in the present momentum convention reads
\begin{equation}
 \bar u(p')i\sigma^{\mu\nu}q_\nu u(p)
 =2\bar p^\mu\mathcal U-2M J_N^\mu.
 \label{eq:gordonq}
\end{equation}
To make the tensor reduction explicit, we consider one representative contraction in each symmetry sector.  For the symmetric tensor, contraction with $\bar p_\mu J_{N\nu}$ reduces the three terms in Eq.~\eqref{eq:symtensortraceless} separately as
\begin{align}
 \bar k^\mu\bar k^\nu\bar p_\mu J_{N\nu}
 &= (\bar k\cdot\bar p)\mathcal K,
 \label{eq:symcontractterm1}\\
 q^\mu q^\nu\bar p_\mu J_{N\nu}
 &= (\bar p\cdot q)(q\cdot J_N)=0,
 \label{eq:symcontractterm2}\\
 g^{\mu\nu}\bar p_\mu J_{N\nu}
 &= \bar p\cdot J_N=M\mathcal U.
 \label{eq:symcontractterm3}
\end{align}
Substitution in Eq.~\eqref{eq:symtensortraceless} then gives
\begin{equation}
 \left(\bar\Phi'T_S^{\mu\nu}\Phi\right)
 \bar p_\mu J_{N\nu}
 =\frac{2\mathcal S}{m^2}\left[
 (\bar k\cdot\bar p)\mathcal K
 -\frac{M}{4}\left(m^2-\frac{t}{2}\right)\mathcal U
 \right].
 \label{eq:symtensorcontract}
\end{equation}
For the antisymmetric tensor, the natural fermion bilinear is $\bar u(p')\sigma_{\mu\nu}u(p)$.  Contracting the antisymmetric DKP matrix element in Eq.~\eqref{eq:antitensor} with the Dirac tensor $\bar u(p')\sigma_{\mu\nu}u(p)$ and using the antisymmetry of $\sigma_{\mu\nu}$ gives the intermediate form
\begin{equation}
 \left(\bar\Phi'T_A^{\mu\nu}\Phi\right)
 \bar u(p')\sigma_{\mu\nu}u(p)
 =-\frac{2\mathcal S}{m^2}\bar k_\mu
 \bar u(p')i\sigma^{\mu\nu}q_\nu u(p).
 \label{eq:antitensorintermediate}
\end{equation}
Applying Eq.~\eqref{eq:gordonq}, together with $\bar k\cdot J_N=\mathcal K$, therefore yields
\begin{equation}
 \left(\bar\Phi'T_A^{\mu\nu}\Phi\right)
 \bar u(p')\sigma_{\mu\nu}u(p)
 =\frac{4\mathcal S}{m^2}\left[
 M\mathcal K-(\bar k\cdot\bar p)\mathcal U
 \right].
 \label{eq:antitensorcontract}
\end{equation}
These two representative contractions illustrate the reduction explicitly; the conclusion does not depend on these particular choices.  The general reduction follows from the allowed parity-even rank-two fermion structures described below.  The most general parity-even symmetric rank-two contractions can be built from the metric, products of the external momenta, and symmetrized products of a momentum with $J_N^\mu$, while the antisymmetric sector can be built from antisymmetrized momentum products, momentum-current products, and $\sigma^{\mu\nu}$.  Equations~\eqref{eq:orthogonality}, \eqref{eq:currentcons}, and \eqref{eq:gordonq} reduce every such on-shell contraction to $\mathcal U$ and $\mathcal K$.

Dual pseudotensor forms do not provide an additional structure because $\sigma^{\mu\nu}\gamma_5=(i/2)\epsilon^{\mu\nu\alpha\beta}\sigma_{\alpha\beta}$.

The complete local spin-zero DKP basis therefore projects, between free elastic on-shell states, onto the same two physical directions already displayed in Eq.~\eqref{eq:AB}.  This statement concerns on-shell matrix elements and their Lorentz contractions; it does not identify the underlying DKP operators.  In particular, the scalar proportionality is momentum-transfer dependent, and the genuine nonminimal vector disappears only in the conserved vector-current channel considered above.

\section{Implications for the KDP relativistic impulse approximation}
\label{sec:ria-implications}

The on-shell reduction derived in Sec.~\ref{sec:onshell-reduction} can be applied directly to the scalar--vector choices used in the traditional KDP relativistic impulse approximation (RIA)~\cite{PRL55:592:1985,PTP103:321:2000}. Following the convention of Refs.~\cite{PRL55:592:1985,PTP103:321:2000}, Case 1 is defined by choosing $I_NP$ and $\gamma_\mu P\beta^\mu$ as the scalar and projected-vector covariants. The corresponding elementary meson--nucleon $t$ matrix is given by
\begin{equation}
 \hat t^{(1)}(s,t)
 =t_s^{(1)}(s,t)\,I_NP
 +t_v^{(1)}(s,t)\,\gamma_\mu P\beta^\mu .
 \label{eq:tcase1}
\end{equation}
Here $I_N$ is the identity in the nucleon Dirac space, and $t_s^{(1)}$ and $t_v^{(1)}$ are the invariant functions associated with the scalar and projected-vector covariants, respectively. To match Eq.~\eqref{eq:tcase1} to the Pauli amplitude in Eq.~\eqref{eq:fg}, we reduce its scalar and vector terms separately in the same Pauli basis.

Using the scalar reduction in Eq.~\eqref{eq:scalarfermionclosure}, the scalar term of Eq.~\eqref{eq:tcase1} becomes
\begin{equation}
 \mathcal M_s^{(1)}
 =\big[\bar u(p')u(p)\big]
  \big[\bar\Phi(k')P\Phi(k)\big]
 =\mathcal S\,\bar u(p')u(p).
 \label{eq:case1scalarfactor}
\end{equation}
The free DKP state used in this matching is normalized by $\bar\Phi(k)\beta^\mu\Phi(k)=k^\mu/m$. At equal external momenta, Eq.~\eqref{eq:vectorreduction} then gives $\bar\Phi(k)P\Phi(k)=1/2$. For the standard five-component spin-zero free state used for this matching, $P$ selects the momentum-independent scalar component, so the same normalization gives
\begin{equation}
 \mathcal S=\bar\Phi(k')P\Phi(k)=\frac12 .
 \label{eq:Shistorical}
\end{equation}

With Eq.~\eqref{eq:Shistorical}, the scalar matrix element in Eq.~\eqref{eq:case1scalarfactor} reduces to $\mathcal M_s^{(1)}=\frac{1}{2}\bar u(p')u(p)$. To express this matrix element in the Pauli basis, we use the center-of-mass Dirac spinor
\begin{equation}
 u(p)=\sqrt{\frac{E+M}{2M}}
 \begin{pmatrix}
  \chi\\[1mm]
  \dfrac{\bm\sigma\cdot\bm p}{E+M}\chi
 \end{pmatrix},
 \label{eq:CMDiracspinor}
\end{equation}
where $\chi$ is a two-component Pauli spinor and $E=\sqrt{M^2+|\bm p|^2}$. Substituting the center-of-mass spinors of Eq.~\eqref{eq:CMDiracspinor} into $\frac{1}{2}\bar u(p')u(p)$ gives
\begin{equation}
 \mathcal M_s^{(1)}
 =\frac{1}{4M(E+M)}\,
 \chi'^\dagger
 \left[(E+M)^2
 -(\bm\sigma\cdot\bm p')(\bm\sigma\cdot\bm p)\right]\chi.
 \label{eq:case1scalarspinor}
\end{equation}
Using $(\bm\sigma\cdot\bm a)(\bm\sigma\cdot\bm b)=\bm a\cdot\bm b+i\bm\sigma\cdot(\bm a\times\bm b)$, together with $\bm p'\cdot\bm p=K_{\rm cm}^2\cos\theta$ and $\bm p'\times\bm p=-K_{\rm cm}^2\sin\theta\,\hat{\bm n}$, where $K_{\rm cm}=|\bm p|=|\bm p'|$ and $\theta$ is the center-of-mass scattering angle, Eq.~\eqref{eq:case1scalarspinor} takes the Pauli form
\begin{equation}
 \mathcal M_s^{(1)}
 =K_{11}^{(1)}
 +K_{21}^{(1)}\,\bm\sigma\cdot\hat{\bm n},
 \label{eq:case1scalardecomp}
\end{equation}
with
\begin{align}
 K_{11}^{(1)}
 &=
 \frac{(E+M)^2-\kappa^2+Q^2/4}
 {4M(E+M)},
 \label{eq:K11}\\
 K_{21}^{(1)}
 &=
 \frac{i\kappa Q}{4M(E+M)}.
 \label{eq:K21}
\end{align}
Here $Q=2K_{\rm cm}\sin(\theta/2)$ and $\kappa=K_{\rm cm}\cos(\theta/2)$; elastic center-of-mass kinematics gives $t=-Q^2$.

For the projected-vector term in Eq.~\eqref{eq:tcase1}, Eqs.~\eqref{eq:nmzero} and \eqref{eq:projectedequiv} give directly
\begin{equation}
 \begin{aligned}
 \mathcal M_v^{(1)}
 &=J_{N\mu}\,\bar\Phi(k')P\beta^\mu\Phi(k)
 =\frac{\mathcal S}{m}\,\mathcal K\\
 &=\frac{\mathcal S}{m}\,\bar u(p')\!\not\!\bar k\,u(p)
 =\frac{1}{2m}\bar u(p')\!\not\!\bar k\,u(p),
 \end{aligned}
 \label{eq:case1vectorME}
\end{equation}
where the last equality uses Eq.~\eqref{eq:Shistorical}. Evaluating the Dirac bilinear in Eq.~\eqref{eq:case1vectorME} with the center-of-mass spinor in Eq.~\eqref{eq:CMDiracspinor} gives
\begin{equation}
 \begin{aligned}
 \mathcal M_v^{(1)}
 =\frac{E+M}{4mM}\,\chi'^\dagger\Bigg[
 &\bar k^0
 -\frac{(\bm\sigma\cdot\bar{\bm k})(\bm\sigma\cdot\bm p)
 +(\bm\sigma\cdot\bm p')(\bm\sigma\cdot\bar{\bm k})}{E+M}\\
 &+\frac{\bar k^0(\bm\sigma\cdot\bm p')(\bm\sigma\cdot\bm p)}{(E+M)^2}
 \Bigg]\chi .
 \end{aligned}
 \label{eq:case1vectorspinor}
\end{equation}
The three Pauli products in Eq.~\eqref{eq:case1vectorspinor} are reduced with the same Pauli identity. In the center-of-mass frame, $\bar{\bm k}\cdot\bm p=\bm p'\cdot\bar{\bm k}=-\kappa^2$ and $\bm p'\cdot\bm p=\kappa^2-Q^2/4$. The corresponding cross products are $\bar{\bm k}\times\bm p=\bm p'\times\bar{\bm k}=(\kappa Q/2)\hat{\bm n}$ and $\bm p'\times\bm p=-\kappa Q\hat{\bm n}$. Substituting these center-of-mass relations into Eq.~\eqref{eq:case1vectorspinor} gives the Pauli decomposition
\begin{equation}
 \mathcal M_v^{(1)}
 =K_{12}^{(1)}
 +K_{22}^{(1)}\,\bm\sigma\cdot\hat{\bm n},
 \label{eq:case1vectordecomp}
\end{equation}
with
\begin{align}
 K_{12}^{(1)}
 &=
 \frac{E_m}{m}\,
 \frac{(E+M)^2+\kappa^2-Q^2/4}
 {4M(E+M)}
 +\frac{\kappa^2}{2mM},
 \label{eq:K12}\\
 K_{22}^{(1)}
 &=
 -\frac{i\kappa Q}{4M(E+M)}\frac{E_m}{m}
 -\frac{i\kappa Q}{4mM}.
 \label{eq:K22}
\end{align}
Here $E_m=\bar k^0=\sqrt{m^2+K_{\rm cm}^2}$ is the meson center-of-mass energy.

The scalar coefficients in Eqs.~\eqref{eq:K11} and \eqref{eq:K21} and the projected-vector coefficients in Eqs.~\eqref{eq:K12} and \eqref{eq:K22} define the Case~1 matching matrix
\begin{equation}
 \mathsf K_1=
 \begin{pmatrix}
 K_{11}^{(1)} & K_{12}^{(1)}\\
 K_{21}^{(1)} & K_{22}^{(1)}
 \end{pmatrix}.
 \label{eq:K1def}
\end{equation}
The first and second columns contain the scalar and projected-vector contributions, respectively, while the first and second rows determine the spin-nonflip amplitude $f$ and the spin-flip amplitude $g$. The Case~1 matching is therefore
\begin{equation}
 \begin{pmatrix} f\\ g\end{pmatrix}
 =\mathcal N\,\mathsf K_1
 \begin{pmatrix}t_s^{(1)}\\ t_v^{(1)}\end{pmatrix}.
 \label{eq:Kgeneric}
\end{equation}
Here $\mathcal N=-Mm/(2\pi\sqrt{s})$ is the normalization factor used in Refs.~\cite{PRL55:592:1985,PTP103:321:2000}.

In the same convention, Case~2 uses $I_NP$ and $\gamma_\mu\beta^\mu$, Case~3 uses $I_N\I$ and $\gamma_\mu P\beta^\mu$, and Case~4 uses $I_N\I$ and $\gamma_\mu\beta^\mu$ as the scalar and vector covariants, respectively. The on-shell relations in Sec.~\ref{sec:onshell-reduction} give a factor of two between the minimal and projected-vector contributions and a factor $2c(Q)$ between the identity and $P$ scalar contributions. Consequently,
\begin{align}
 \mathsf K_2
 &=\mathsf K_1
 \begin{pmatrix}1&0\\0&2\end{pmatrix},
 \label{eq:K2}\\
 \mathsf K_3
 &=\mathsf K_1
 \begin{pmatrix}
 2c(Q)&0\\
 0&1
 \end{pmatrix},
 \label{eq:K3}\\
 \mathsf K_4
 &=\mathsf K_1
 \begin{pmatrix}
 2c(Q)&0\\
 0&2
 \end{pmatrix}.
 \label{eq:K4}
\end{align}
where $c(Q)=1+Q^2/(4m^2)$.
Since Cases~1--4 reproduce the same Pauli amplitudes, Eqs.~\eqref{eq:K2}--\eqref{eq:K4} imply that the corresponding invariant functions satisfy
\begin{align}
 t_s^{(2)}&=t_s^{(1)},
 &\qquad t_v^{(2)}&=\frac12 t_v^{(1)},
 \label{eq:t2}\\
 t_s^{(3)}&=\frac{t_s^{(1)}}{2c(Q)},
 &\qquad t_v^{(3)}&=t_v^{(1)},
 \label{eq:t3}\\
 t_s^{(4)}&=\frac{t_s^{(1)}}{2c(Q)},
 &\qquad t_v^{(4)}&=\frac12 t_v^{(1)}.
 \label{eq:t4}
\end{align}
Thus Cases~1--4 represent the same two physical on-shell directions through different DKP covariants; the case labels do not introduce additional elementary information.

Unlike the vector representatives used in Cases~1--4, the genuine nonminimal vector $i[P,\beta^\mu]$ does not contribute to the conserved on-shell vector-current channel. From Eq.~\eqref{eq:nmzero}, the contraction $J_{N\mu}\bar\Phi' i[P,\beta^\mu]\Phi$ vanishes identically. Using $i[P,\beta^\mu]$ as the vector covariant therefore makes the vector column of $\mathsf K$ vanish, so the matching matrix is singular and the corresponding invariant function cannot be determined from the free on-shell amplitudes $f$ and $g$.

The preceding discussion concerns only elementary on-shell matching. The next question is whether equivalent representatives remain equivalent after they are folded with nuclear densities and inserted into the DKP optical equation. We first consider the vector embedding. Following the RIA construction of Refs.~\cite{PLB427:231:1998,PTP103:321:2000}, the vector optical potential for Case~$a$ is
\begin{equation}
 U_v^{(a)}(\mathbf r)
 =-\frac{P_{\rm lab}}{(2\pi)^2m}
 \sum_{N=p,n}\int\frac{d^3q}{(2\pi)^3}\,
 e^{i\mathbf q\cdot\mathbf r}\,
 t_{v,N}^{(a)}(q)\,\widetilde\rho_{v,N}(q).
 \label{eq:Uvfolding}
\end{equation}
Here $a=1,\ldots,4$ labels the case, $N=p,n$ labels the proton or neutron channel, $P_{\rm lab}$ is the incident meson laboratory momentum, and $\widetilde\rho_{v,N}$ is the Fourier-transformed nuclear vector density. Since Eq.~\eqref{eq:t2} gives $t_{v,N}^{(1)}=2t_{v,N}^{(2)}$ in each nucleon channel, linearity of the folding in Eq.~\eqref{eq:Uvfolding} gives
\begin{equation}
 U_v^{(1)}=2U_v^{(2)}.
 \label{eq:Uv12}
\end{equation}
After elimination of the auxiliary DKP components, the projected-vector coupling of Case~1 and the minimal-vector coupling of Case~2 lead to different second-order equations. Writing $X=E-U_C$, $U_v^{(1)}=2V$, and $U_v^{(2)}=V$, with $U_C(\mathbf r)$ denoting the static Coulomb potential obtained from the empirical nuclear charge distribution~\cite{PTP103:321:2000}, the corresponding vector-dependent terms are, respectively,
\begin{align}
 X(X-2V)&=X^2-2XV,
 \label{eq:case1vector}\\
 (X-V)^2&=X^2-2XV+V^2.
 \label{eq:case2vector}
\end{align}
The two optical operators therefore agree through first order in the vector potential and first differ at quadratic order,
\begin{equation}
 \Delta\mathcal E_v=V^2.
 \label{eq:vectorquadratic}
\end{equation}
The same quadratic difference also arises between Cases~3 and 4, since these two cases share the same scalar embedding and differ only by the replacement of the projected-vector coupling by the minimal-vector coupling.
Thus the vector-sector distinction between the projected and minimal couplings arises from the nonlinear optical embedding, even though they represent the same elementary vector content on shell.

The scalar Cases~1/2 and 3/4 exhibit the same principle in a less trivial form because their on-shell proportionality factor depends on $Q^2$. In direct analogy with Eq.~\eqref{eq:Uvfolding}, the scalar optical potential for Case~$a$ is
\begin{equation}
 U_s^{(a)}(\mathbf r)
 =-\frac{P_{\rm lab}}{(2\pi)^2m}
 \sum_{N=p,n}\int\frac{d^3q}{(2\pi)^3}\,
 e^{i\mathbf q\cdot\mathbf r}\,
 t_{s,N}^{(a)}(q)\,\widetilde\rho_{s,N}(q).
 \label{eq:Usfolding}
\end{equation}
Here $\widetilde\rho_{s,N}$ is the Fourier-transformed nuclear scalar density; the remaining symbols have the same meaning as in Eq.~\eqref{eq:Uvfolding}. Let $U_P(Q)$ denote the momentum-space scalar potential associated with the projected scalar $P$ in Cases~1/2 and $U_I(Q)$ that associated with $\I$ in Cases~3/4. The scalar invariant relations in Eqs.~\eqref{eq:t2}--\eqref{eq:t4}, together with the linearity of the folding, give
\begin{equation}
 U_P(Q)
 =
 2\left(1+\frac{Q^2}{4m^2}\right)U_I(Q).
 \label{eq:UPUIq}
\end{equation}
Using $Q^2\leftrightarrow-\nabla^2$ in coordinate space, Eq.~\eqref{eq:UPUIq} becomes
\begin{equation}
 \left(1-\frac{\nabla^2}{4m^2}\right)U_I(\mathbf r)
 =
 \frac12 U_P(\mathbf r).
 \label{eq:helmholtzscalar}
\end{equation}
The scalar equivalence therefore becomes a differential relation after folding rather than a constant rescaling.

The consequence of this momentum-dependent scalar relation becomes explicit after elimination of the auxiliary DKP components. To isolate the scalar embedding, we compare cases with the same vector coupling. For the projected-vector coupling, Cases~1 and 3 give the second-order operators
\begin{align}
 \mathcal O_1
 &=X(X-U_v)-m\bigl(m+U_P\bigr)+\nabla^2,
 \label{eq:case1scalarop}\\
 \mathcal O_3
 &=X(X-U_v)-\mu^2
 +\mu\,\boldsymbol\nabla\!\cdot
 \left(\mu^{-1}\boldsymbol\nabla\right).
 \label{eq:case3scalarop}
\end{align}
where $\mu(\mathbf r)=m+U_I(\mathbf r)$. For the minimal-vector coupling, Cases~2 and 4 give
\begin{align}
 \mathcal O_2
 &=(X-U_v)^2-m\bigl(m+U_P\bigr)+\nabla^2,
 \label{eq:case2scalarop}\\
 \mathcal O_4
 &=(X-U_v)^2-\mu^2
 +\mu\,\boldsymbol\nabla\!\cdot
 \left(\mu^{-1}\boldsymbol\nabla\right).
 \label{eq:case4scalarop}
\end{align}
Within each pair the vector-dependent terms are identical, so the comparison isolates the scalar embedding. The projected scalar contributes $-mU_P$, whereas the identity scalar modifies both the mass and derivative terms through $\mu=m+U_I$. The differential operator can be written as
\begin{equation}
 \mu\,\boldsymbol\nabla\!\cdot
 \left(\mu^{-1}\boldsymbol\nabla\right)
 =
 \nabla^2
 -\frac{\boldsymbol\nabla\mu}{\mu}\cdot\boldsymbol\nabla.
 \label{eq:mudifferential}
\end{equation}
Subtracting the free contribution $-m^2+\nabla^2$ and expanding to first order in $U_I$ gives
\begin{equation}
 \delta\mathcal O_I
 =
 -2mU_I
 -\frac{1}{m}
 (\boldsymbol\nabla U_I)\cdot\boldsymbol\nabla.
 \label{eq:deltaOI}
\end{equation}
Evaluating Eq.~\eqref{eq:deltaOI} between initial and final elastic plane-wave states converts the derivative term into its momentum-space form, giving
\begin{equation}
 \langle\delta\mathcal O_I\rangle
 =
 -2m\left(1+\frac{Q^2}{4m^2}\right)U_I(Q)
 =
 -mU_P(Q),
 \label{eq:scalarfirstorder}
\end{equation}
where Eq.~\eqref{eq:UPUIq} was used in the last step. Thus Cases~1/3 and Cases~2/4 reproduce the same elementary first-order on-shell scalar content. Beyond first order, however, the identity embedding generates both powers of $U_I$ and the derivative structure in Eq.~\eqref{eq:mudifferential}, so the corresponding optical operators need not remain identical.

The historical case dependence can therefore be separated into two logically distinct statements. At the elementary level, the four scalar--vector choices are different representatives of the same two-dimensional on-shell amplitude space established in Secs.~\ref{sec:scattering-covariants} and \ref{sec:onshell-reduction}. Differences between their nuclear predictions arise only after those representatives are embedded and iterated as optical operators, where quadratic and derivative terms retain information about the chosen representation.

\section{Discussion and conclusions}
\label{sec:discussion}

In this work, we examined the on-shell content of the scalar, vector, and tensor DKP covariants relevant to the spin-zero relativistic impulse approximation. We first reduced their free on-shell matrix elements to the physical spin-zero--spin-$1/2$ amplitude space and then applied this reduction to the scalar--vector choices used in the historical KDP-RIA Cases~1--4~\cite{PRL55:592:1985,PTP103:321:2000}. The first-order meson--nucleon input therefore contains only the two physical amplitudes allowed by on-shell spin-zero--spin-$1/2$ scattering. The larger set of DKP matrices does not introduce additional elementary observables; rather, different covariants provide different representations of the same two physical directions.

The on-shell reduction makes this statement explicit across the DKP basis. The scalar covariants collapse onto a single scalar direction, while the vector covariants contain one physical transverse direction after contraction with the conserved fermion current. In particular, the genuine nonminimal vector $i[P,\beta^\mu]$ is longitudinal and its contraction with the conserved fermion current vanishes, whereas the projected vectors $P\beta^\mu$ and $\beta^\mu P$ each reproduce one half of the minimal-vector contribution in this channel. The tensor covariants likewise introduce no additional on-shell directions: after contraction with the fermion bilinears, their contributions reduce to the same two Dirac structures that span the physical amplitude.

Applied to the historical scalar--vector choices~\cite{PRL55:592:1985,PTP103:321:2000}, this reduction shows that Cases~1--4 are different representations of the same two-dimensional elementary on-shell amplitude space. Their invariant functions therefore differ only by the rescalings required to compensate for the chosen DKP representatives and to reproduce the same Pauli amplitudes $f$ and $g$. The case labels consequently do not encode additional elementary meson--nucleon information.

The distinction among the cases emerges only after these equivalent representatives are embedded into the nuclear optical equation. For the projected-versus-minimal vector comparison, the first-order contribution is unchanged and the first difference appears at quadratic order in the folded vector potential. For the projected-versus-identity scalar comparison, the two embeddings likewise reproduce the same first-order on-shell contribution, while the identity embedding generates both quadratic terms in $U_I$ and the derivative structure associated with its second-order DKP operator. Thus the elementary on-shell equivalence of the covariants does not imply identical optical operators beyond first order.

This separation between elementary on-shell content and optical embedding provides a direct criterion for interpreting phenomenological comparisons among different KDP-RIA implementations. Differences among Cases~1--4 in meson--nucleus observables should therefore be traced to the nonlinear and derivative structures generated by the optical embedding rather than to additional elementary amplitudes. A natural next step is to quantify these effects in pion- and kaon-nucleus scattering~\cite{PLB427:231:1998,PTP103:321:2000} and to determine the kinematic regimes in which the different embeddings become phenomenologically distinguishable.

\section*{Funding}
This work was supported in part by means of funds provided by CNPq, Brazil, Grant Nos. 308172/2023-0 and CNPq/Universal/420896/2025-2, FAPEMA, and CAPES (Finance Code 001).

\section*{Declaration of competing interest}
The authors declare that they have no known competing financial interests or personal relationships that could have appeared to influence the work reported in this paper.


\section*{Data availability}
No data was used for the research described in the article.




\bibliographystyle{elsarticle-harv} 
\bibliography{mybibfile2020}






\end{document}